# ASSESSMENT OF URBAN ROOFTOP GRID CONNECTED SOLAR POTENTIAL IN NEPAL – A CASE STUDY OF RESIDENTIAL BUILDINGS IN KATHMANDU, POKHARA AND BIRATNAGAR CITIES

J. N. Shrestha*, D. B. Raut**

*Center for Energy Studies, Institute of Engineering, Tribhuvan University
Pulchowk, Lalitpur, Nepal
shresthajn@gmail.com
** Thapathali Campus, Institute of Engineering, Tribhuvan University, Kathmandu, Nepal
raut.debendra@tcioe.edu.np

ABSTRACT

This paper assesses the technical, financial, and market potential of the rooftop Solar Photovoltaic (PV) system on residential buildings in major cities namely Kathmandu valley, Pokhara, and Biratnagar of Nepal. Three sets of questionnaires were prepared each for residential households, PV suppliers, and solar project financing institutions. From the field survey, it is found that the average rooftop area available for PV installation is 14.5 sq.m, 12.45 sq.m, and 19 sq.m in Kathmandu, Pokhara, and Biratnagar cities respectively. Considering 557,027 residential buildings in Kathmandu; 77,523 in Pokhara and 33,075 in Biratnagar, total rooftop PV power potential in all three cities are found to be 970 MWp which could generate 1,310 GWh/year that comes out to be 35% of the electricity sold by Nepal Electricity Authority (NEA) in fiscal year 2014/15. Based on the 1.5 kWp PV system design per household and market price of 2016, the Levelized Cost of Electricity (LCOE) changes from NRs 8/kWh to NRs 20/kWh for basic lighting to a full load consisting of domestic electrical appliances. The technical barriers for the grid connection of rooftop solar in Nepal are not a major issue now as Nepal Electricity Authority has set clear guidelines for it.

## INTRODUCTION

To resolve the existing severe power crisis and enhance energy security by diversifying Nepal's energy mix, abundant solar PV potential needs to be harnessed. Therefore, the need for promoting solar PV energy has been rightly emphasized in the recent report by the Ministry of Energy, Government of Nepal entitled "National Energy Crisis Mitigation 2016". A study (Nepal, 2014) shows that about 300,000 residential buildings in Kathmandu valley itself are using battery-inverter backup systems that consume heavily subsidized cheap grid electricity during peak demand times and drastically increases the pressure on the national grid, and hence impose more pressure on the current supply system and exacerbates the already severe load shedding.

Although the efficiency of the electrical inverter backup system is low, users are compelled to use this due to the high initial investment for solar energy (Pokharel, 2016). At present, in major cities of Nepal such as the Kathmandu valley, rooftop solar PV systems have not been broadly installed yet either by residential or by commercial and industrial users. A better understanding of the market from the demand, supply, and regulatory aspects is urgently needed before decisions on proper intervention can be made (Nakarmi, 2014).
A similar study (Chanese et al., 2009) was carried out in Kathmandu valley found that an average of 25 m2 available roof surfaces for PV and about 5 % roof space lies between 50 to 100 m2. But this study was conducted in Kathmandu valley only and didn't estimate the total rooftop potential, energy generation potential, critical aspects from financing, and suppliers were not conducted in this study.

The overall objective of this paper is to assess the potential of rooftop solar PV in major cities of Nepal namely Kathmandu valley (Ktm.), Pokhara (Pkr.) and Biratnagar (Birt.). This paper also explores the barriers to further development of rooftop PV and recommends proper interventions required to address the barriers. To meet the principal objectives, the analysis was done from the demand, supply, financing, and regulatory point of view. The demand-side analysis explored the consumer's electricity consumption pattern and rooftop PV installation capacity on residential buildings whereas the supply side accessed the key data from the financing institutions, solar manufacturers, and suppliers.

**METHODOLOGY ADOPTED**

**Data Collection**

Primary data was collected from baseline surveys on the demand side, supply-side and financial institutions that are financing solar PV business in Nepal. Information collected as part of the baseline survey on demand-side includes the demographic and building characteristics, current electricity usage pattern, customer awareness and interest, availability of PV modules in the current market, reasons for customer purchase or non-purchase, willingness to pay, etc. Survey engineers have visited the buildings and measured the detail parameters including the snaps of sampled of the rooftop.

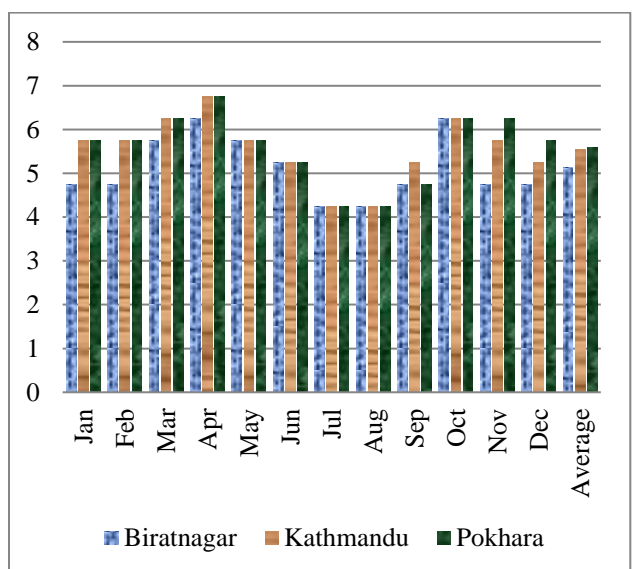

**Figure 1 Solar insolation at flat plate tilted to latitude, kWh/m2/day**

Information collected on supply-side survey includes the records on the supply of PV modules, prices, and market share of the major suppliers, supplier's offer to the consumers in "back-up" mode or only "feed-in" mode, services (pre-and-post sales) provided by the supplier, financing methods and terms of conditions, awareness of government subsidy programs, fiscal incentives such as waiving of VAT and import duties, and any regulations on quality assurance and quality control (NEPQA 2015 revised version) standards, major barriers to rooftop solar PV installation, etc.

Also, secondary data about the utility consumers, total numbers of buildings, etc. were collected from the NEA (Nepal Electricity Authority) and Nepal Census Report 2011.

Similarly, meteorological data for the study area were taken from the Solar and Wind Energy Resource Assessment (SWERA) report. From Fig. 1, the annual average peaksun hours used to estimate the energy in this study are 5.54, 5.58, and 5.13 for Kathmandu, Pokhara, and Biratnagar city respectively (Schillings et al., 2004).

**Sampling**

Under the predefined sample size, the random samples were selected based on the economic status, classes, and power consumption profile.

**Population**

From Nepal census 2011 data (CBS 2011), total numbers of residential buildings are projected to 2015 with growth rates of 6%, 12% and 4% per year in Kathmandu, Pokhara and Biratnagar cities respectively. Therefore, the total population considered in this study is 557,027; 77,523 and 33,075 for Kathmandu, Pokhara and Biratnagar cities respectively. Residential buildings considered are only cement bonded bricks/stone and with RCC structure only and buildings with mud bonded bricks/stone and wooden pillar are excluded in all the cities.

**Data Analysis**

Statistical analysis tools (Isreal, 1998) is used with a precision level (e)= ±7% and 95% of confidence level and P=0.5. Data collected from residential buildings are classified with the footprint area as small (<100 m2), medium (100m2 -150m2), and large (>150 m2). The sample means and standard deviation of each category area calculated as E and $\sigma$ respectively. The statistical tools used to determine the population mean and extrapolate the total rooftop solar potential of the whole population is presented in Table 1.

Table 1 Summary of statistical tools used

| | |
|---:|:---|
| Z-test with Confidence Level | **0.95** |
| Value of Z at 95% Confidence Level is (Z) | **1.96** |
| Sample Mean and Population Mean | **E, μ** |
| Sample Standard Deviation | **σ** |
| Then, Population mean is calculated with following relations | $Z = \dfrac{(E-\mu)}{\left(\dfrac{\sigma}{\sqrt{n}}\right)}$ |
| Total estimations | **μ * N** |

**Key Assumptions Considered to Estimate Shade Free Area and PV Potential**

The available shade-free PV installation rooftop area estimation, kWp potential, and final yield from PV estimation are the major technical challenges of rooftop PV study. The assumption and steps carried for these key parameters are presented in Fig. 2 and Fig. 3. Consumers are not willing to give all shade-free rooftops area for PV installation because rooftops are used for other purposes like drying clothes, social and cultural aspects, etc. Therefore, based on the respondent's statements considering all these Nepalese social-cultural aspects, on an average 30% of the shadow-free areas of the rooftop are considered in this study as available space for PV modules in residential sectors, as shown in figure 1.1.

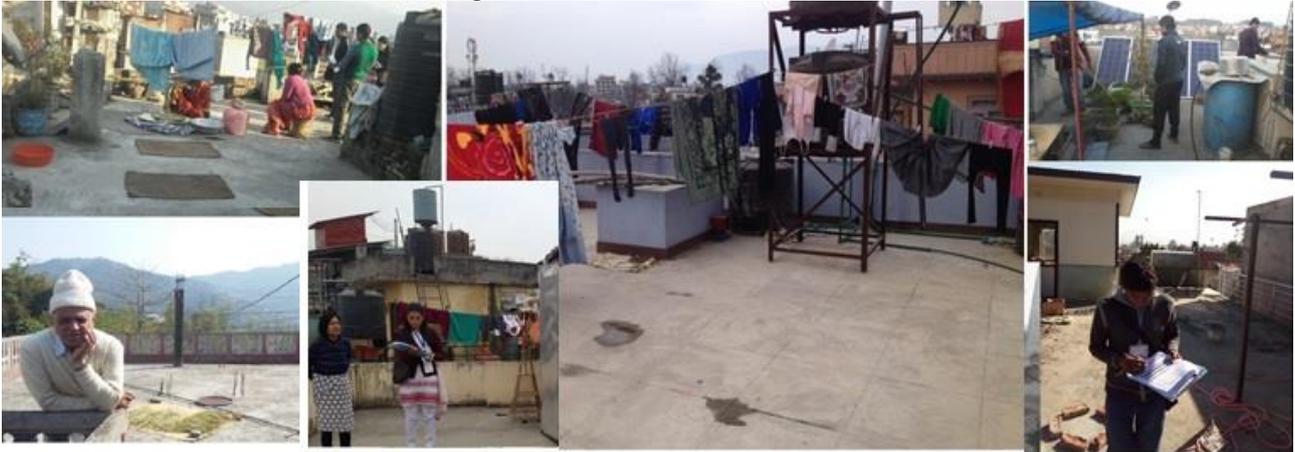

Figure 1.1 Rooftops of sample residential buildings collected during field survey

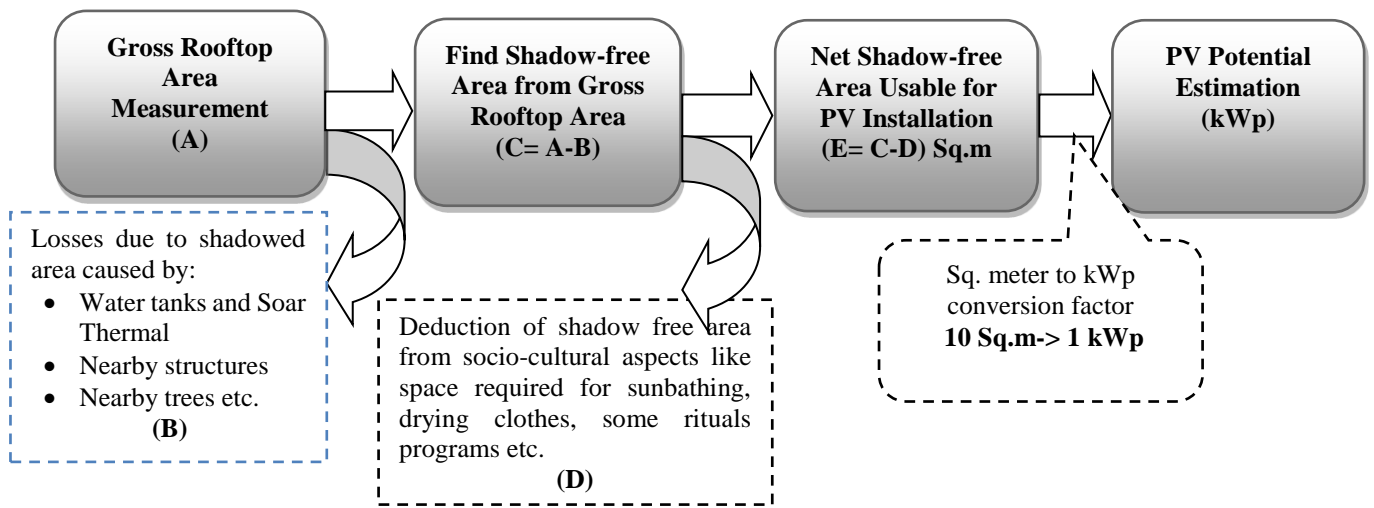

**Figure 2 Shadow free area and PV potential estimation steps for residential buildings**

Energy calculations for example, in reference to Fig. 3, residential buildings in Kathmandu valley will produce 0.67*5.54=3.7 kWh/kWp/day of end-use energy however, it may vary as it depends on the efficiency of BOS used in a real field project.

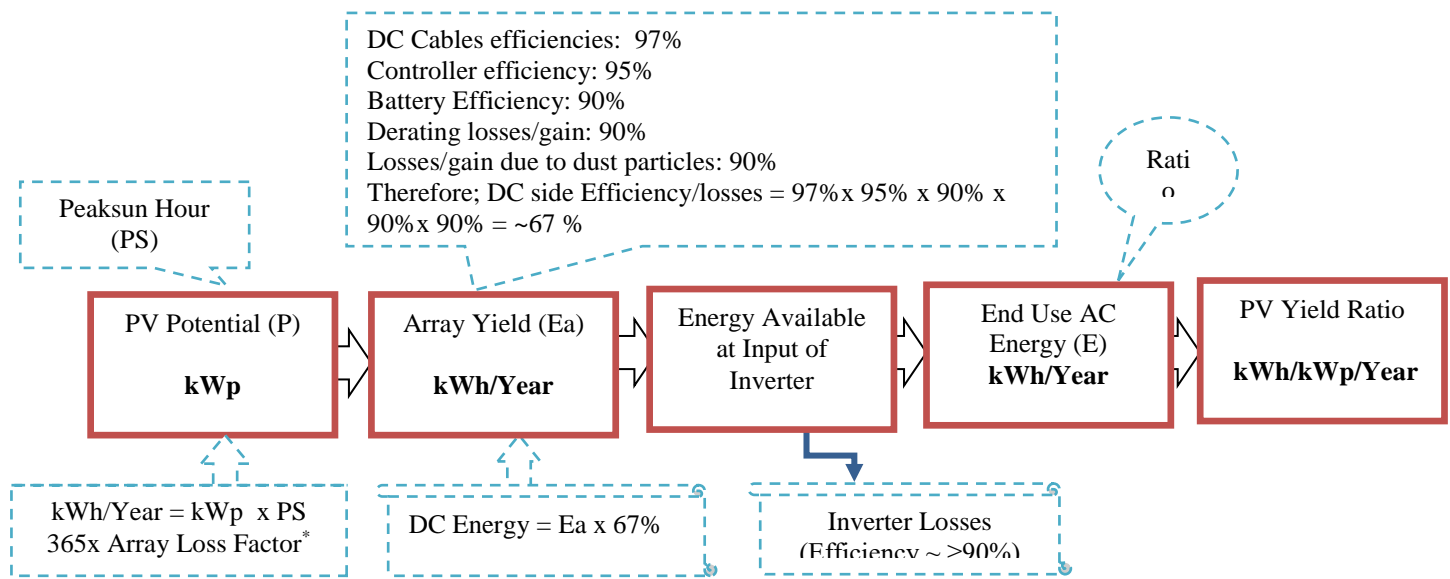

**Figure 3 Steps to determine the energy generation potential from PV power capacity potential**

## RESULTS AND DISCUSSIONS

### A. Existing Power Backup System

It is quite interesting to know that about 35% of the residential buildings in all cities use solar PV systems to power their home during load shedding hours as shown in Fig. 3 (a). In addition, more than 60% of residential buildings use less than 100 Wp solar PV system for lighting purpose during the power outages among which the most common size are 20 Wp and 40 Wp PV modules. However, the most common use of inverter in the inverter-battery-backup mode was 600 VA to 850 VA for Kathmandu and Pokhara but residences in Biratnagar most commonly (80%) used 1400 VA to 1500 VA inverter to power all the basic electrical devices with first a priority to lighting and cooling. In Kathmandu, more than 60% of the PV could backup full hours of load shedding with designed loads and only 5% backups the grid outage partially. Also, it is revealed that more than 40% of the consumers satisfied with their PV systems. Regarding the age of existing PV backup, 22% of the PV systems in Kathmandu and 33% in Biratnagar are more than 5 years old and likely to replace the storage battery while in Pokhara, about 70% of the solar PV system is purchased within last 2 years.

Regarding the source of income, about 80% of the consumers in Kathmandu purchased their PV system from service income whereas more than two-thirds of the consumers in Pokhara and Biratnagar city purchased from their business income. It is to be noted that the number of consumers purchasing solar PV systems with a subsidy is insignificant. Though the government of Nepal provides a subsidy for urban rooftop solar system since 2015, it can be said that almost all consumers in three cities purchased their PV systems without subsidy.

### B. Existing Power Consumption Profile

The study revealed that there are more than 200,000 residential buildings in three cities that have 5A utility meter and use basic electrical appliances only, also about 19% residential buildings in Kathmandu, 40% in Pokhara and less than 5% in Biratnagar cities consume less than 80 units per month, which is a nominal unit set by NEA.

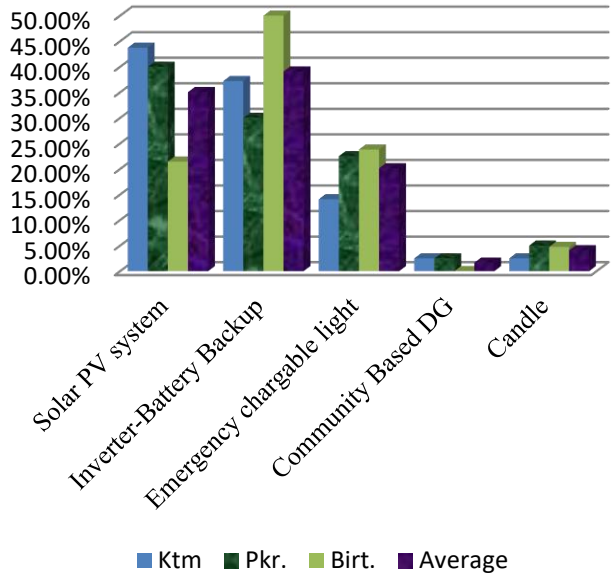
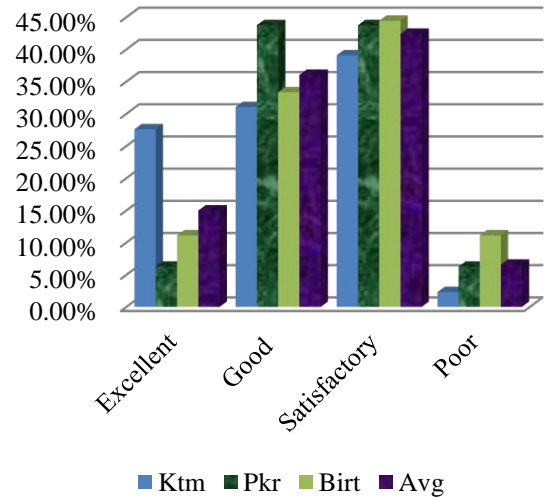

(a)          (b)

**Figure 3 (a) Existing power backup technology used in residential buildings by type (b) consumer satisfactions with existing roof top solar**

### C. Consumer Awarness and Major Barriers

Consumer awareness is the major challenge for adopting and promoting urban rooftop solar program in Nepal. This study shows that more than 80% of the respondents either do not know or partly know about the government subsidy via the "Urban Solar PV Subsidy Program 2015". This factor clearly indicates that an effective awareness campaign is essential immediately. Regarding the awareness of quality assurance, almost all respondents (97%) are not aware of Renewable Energy Test Station (RETS) and Nepal PV Quality Assurance (NEPQA) standards that are key technical guidelines for quality assurance in Nepal.

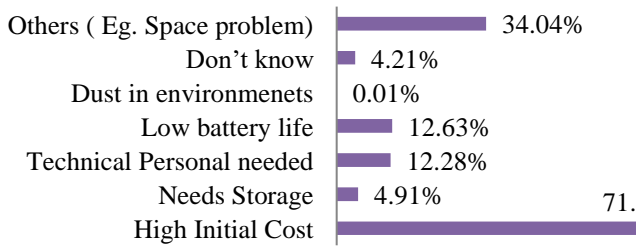

**Figure 4 Consumer awareness on major bottleneck to urban solar program in Nepal**

As can be seen from Fig. 4, most of the respondents are very much concerned with the initial cost of PV systems. They also pointed out that after-sales service and low battery life are hampering widespread use of PV systems. Some respondents expressed their opinion that they are unable to use PV systems as they do not have enough sunny space; quality of BOS, etc.

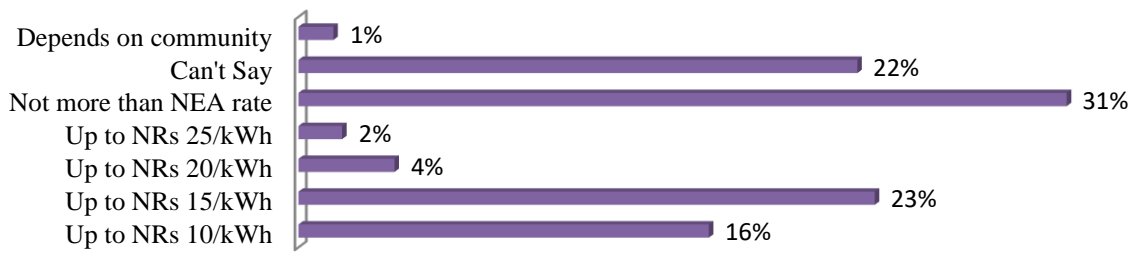

**Figure 5 Consumer response on willingness to pay for 24*7 hour available electricity in all cities in residential sectors**

Fig. 5 show that 23% of the respondents are ready to pay up to Rs. 15/unit should they be provided 24/7 hours of reliable electricity by the utility. This is almost two times higher than the average unit cost of electricity in Nepal. 31% of the respondents are ready to pay as per NEA set tariff for 24/7 hours uninterruptible electricity supply. This clearly indicates the importance of electricity supply in Nepalese society.

**D. Roof top PV Potential and Technical Feasibility**

The estimations of roof top potential includes following:
- Gross roof top area
- Shade free roof top area
- Area available for PV installation
- Estimated kWp potential
- Estimated energy generation (kWh/year) potential

It has been observed that Kathmandu valley has diversified rooftop size and building height ranging from 4 to 20 meters. But in Pokhara city, buildings are uniform in size and height but in Biratnagar, buildings are wider spaced with lower heights.

Statistical data analysis (confidence level of 95% and Z=1.96) shows that the sample means of gross rooftop area of Kathmandu is 80.39 m$^2$ (sampled data varies from 12.55 m$^2$ to 208 m$^2$), it is 72.48 (sample variation 9.29 m$^2$ to 335 m$^2$) for Pokhara and 116.83 m$^2$ (samples from 40 m$^2$ to 552 m$^2$). From this scenario, the population mean is 76.21 m$^2$, 51.25 m$^2$ and 91 m$^2$ for three respective cities with all average of **72.66 m$^2$** of gross rooftop area of residential buildings from all cities. From Table 2, it can be concluded that every residential building has a ~15m$^2$ area available for PV modules. In summary, 72.66 m$^2$ is the gross rooftop area, 51.32 m$^2$ is the shadow-free area and 15.39 m$^2$ is the rooftop area available for PV installation in all cities with total PV potential of 810MWp, 96MWp and 63 MWp in respective cities.

**Table 2 Summary of roof top PV potential in residential sectors**

| Measuring Parameters | Ktm | Pkr | Birt |
|---|---|---|---|
| Total gross roof top area (km$^2$) | 42.45 | 3.97 | 3 |
| Total area available for PV Installation (km$^2$) | 8.10 | 0.97 | 0.64 |
| Average gross roof top area (m$^2$/building) | 76.21 | 51.25 | 91 |
| Average shade free roof top area (m$^2$/building) | 48.48 | 41.5 | 64.0 |
| Average area available for PV installation (m$^2$/building) | 14.54 | 12.45 | 19.2 |
| Average PV potential ( kWp/building) | 1.45 | 1.24 | 1.92 |
| Total estimated PV potential (MWp) | 810 | 96 | 63 |

According to the methodology presented in this report, the calculated annual energy potential (GWh/year) from the rooftop of residential buildings are 1097, 131 and 79 for respective cities totaling 1308 GWh which is 35% of the total utility energy sold which is 3,743 GWh (NEA, 2016) for the country in the fiscal year 2014/15.

**Table 3 Energy consumption patterns of residential building Vs roof top shade-free area**

| Building Class | Total Building | Average Annual Energy Consumption[1] | Day Time Energy Demand Potential (Max)[2] | Dark Hour Energy Demand Potential (Max)[3] | Total Connected Load[4] |
|---|---|---|---|---|---|
| **Shade free roof top area (m²)** | (%) | kWh/year/ building | kWh/day | | kW/building |
| Kathmandu City | | | | | |
| Small (<20) | 33.51% | 1,956 | 2.91 | 5.03 | 3.19 |
| Medium (20-40) | 55.67% | 2,256 | 2.72 | 5.21 | 3.48 |
| Large (>40) | 10.82% | 3,287 | 5.95 | 6.43 | 4.52 |
| **All Buildings (Average)** | | **2,267** | **3.13** | **5.28** | **3.49** |
| Pokhara City | | | | | |
| Small (<20) | 66.67% | 1,612 | 1.72 | 4.19 | 2.06 |
| Medium (20-40) | 14.29% | 2,505 | 5.90 | 11.69 | 4.38 |
| Large (>40) | 19.05% | 1,649 | 3.72 | 12.48 | 3.95 |
| **All Buildings (Average)** | | **1,747** | **2.69** | **6.84** | **2.75** |
| Biratnagar City | | | | | |
| Small (<20) | 19.05% | 2,171 | 3.59 | 9.63 | 3.23 |
| Medium (20-40) | 19.05% | 2,171 | 3.59 | 9.63 | 3.23 |
| Large (>40) | 64.29% | 3,844 | 8.65 | 16.11 | 4.91 |
| **All Buildings (Average)** | | **3,654** | **7** | **15** | **5** |

**E. Supply Side and Regulatory Side Assessment of Roof top PV Market**

The supply-side assessment covers the supply chain of solar PV modules in the market, the supply of solar PV systems in different modes, financing options by the suppliers, major barriers to adopting rooftop urban solar program by supplier's point of view, etc. In the present market context of Nepal, almost 70% of the PV modules are imported from China and the average cost of Chinese modules is about NRs 87/Wp. Indian modules cover about 15 % of the total market share with an average price of NRs 97/Wp. There is a small PV market share of other countries namely Korea, Japan, USA etc in Nepal as shown in Fig. 6.

Taking about the strength and services provided by PV suppliers in the Nepalese market, 50% of them provide free of cost site surveys and energy audit services to the consumers. Also, 80% of suppliers provide free of cost after-sales service for up to 2 years, 10 % for 1 year, and another 10% for 3 years. 90% of suppliers have their own in house team for operation and maintenance, the technical personnel ranges from 2 to 12 people in each company.

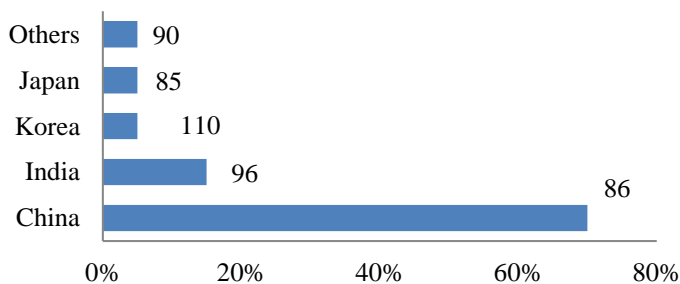

**Figure 6 Imported PV modules by country of origin and average wholesale price (NRs/Wp)**

---

[1] as per utility bill, low consumption due to excessive hours of loadshedding
[2] considering maximum demand if 24*7 electricity is available during day time
[3] considering maximum demand if 24*7 electricity is available during night time
[4] total connected load in the building but all the loads are not used at a time

The financing mechanism is another important aspect of rooftop solar market development. There are five commercial banks authorized to finance under the government subsidy program in Nepal. Out of the five listed banks, only one bank has more than 7 years of experience in solar lending. Without subsidy, it has been charging 12% per annum interest on solar lending but under the subsidy program, all banks lending at 2.25% per annum interest rates on residential use and 4.5% per annum on institutional use considering the proposed solar PV system itself as collateral and validity of loan is 5 years i.e. 60 EMI.

**F. Financial Feasibility of Roof top Solar in Nepal**

For financial feasibility, a 1.5 kWp PV is taken under consideration for this study. Considering the rooftop space, the market potential of 1.5 kWp system is 626 MWp out of which 525 MWp belongs to Kathmandu only, as seen in Fig. 7. However, when energy demand is considered, 1.5 kWp systems best fulfill the demand for average residential buildings in Nepal.

The cost of a 1.5 kWp system in full backup mode comes around NRs 418,032 and it is NRs 322,186 in feed-in mode with partial (40%) storage. To implement on this scenario, total of NRs 174,592million investment is required for backup mode and NRs 134,562 million for feed-in mode under "Urban Rooftop Solar Program 2015".  The LCOE of 1.5 kWp PV in backup mode is calculated to NRs 19.6/kWh years in backup mode and it is NRs 5.89/kWh in feed-in mode, considering the 60 EMI at 2.25% interest and battery replaced every 5 years for 20 years life cycle period. However, the LCOE of the PV system greatly varies from NRs 8/kWh to NRs 20/kWh depending upon the load types to be operated in backup mode, all prices are based on year 2016.

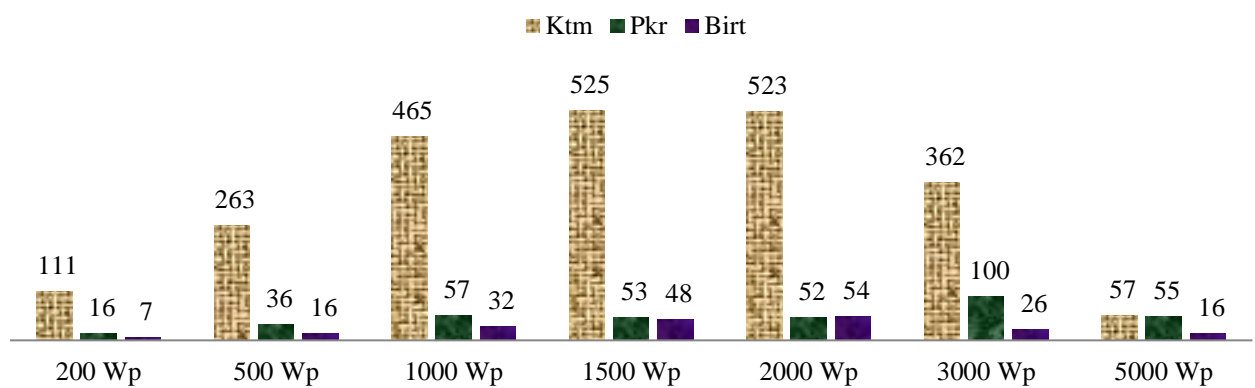

**Figure 7 PV potential in MWp with different system size in three cities**

**G. NEA Requirements for Grid Connection**

Following minimal criteria is set by NEA for grid-connected solar PV (PVGC) system which is approved by NEA Board on 2074/01/13 (26, April 2017):

- Frequency: 50 Hz
- Voltage Level: 230 V/ 400 V/ 11 kV ±5%
- Voltage Waveform: Sinusoidal
- Phase Voltage Imbalance in case of Three Phase System: 1% (maximum)
- Harmonic Distortion (THD): <= 3%
- Power Factor in between: 0.85 Lag and 0.95 Lead
- Energy to be injected to Grid: Not more than 90% of Energy consumed from NEA
- Energy Meter: Bi-directional (Grid Side), Solar Generator Meter (PV Generator Side)

- Power Level Injection: 500 Wp min to up to 5 kWp at 230 V level; more than 5 kWp to up-to 40 kWp at 400 V level; more than 40 kWp at 11 kV level

**CONCLUSIONS**

The gross rooftop area of each residential building was found to be 76m$^2$, 51m$^2$ and 91m$^2$ in Kathmandu, Pokhara and Biratnagar cities respectively totaling 49.42km$^2$ in all cities. After deducting the sunny rooftop space to be used for other purposes, the net area for PV installation comes to be 9.71 km$^2$ which is around 20% of gross rooftop area. Total PV potential in all cities is 970 MWp out of which 810 MWp belongs to Kathmandu Valley alone. If all the residential buildings totaling 667,525 numbers in three cities installed rooftop PV in their sunny space, 35% of the NEA electric energy sold in fiscal year 2014/15 could be saved annually. The average cost of 1.5kWp PV systems is estimated as NRs 418,032 in backup mode and NRs 322,186 in feed-in mode and LCOE vary from NRs 8- NRs 20 per kWh depending upon the operating loads.

There is a good market potential of rooftop solar in Nepal as NEA has set the guidelines for grid-connected PV and there is no load shedding in the day time in major cities. However, a clear financing mechanism and implementation modality are the major challenges for the promotion of the rooftop solar system in Nepal.

**ACKNOWLEDGMENT**

This study was supported by AEPC. The authors wish to acknowledge the following for their valuable contributions in this study: Dr. Shree Raj Shakya, Dr. Mangala Shrestha, AEPC officials Mr. R. P. Dhital, Mr. S. Gautam, Mr. M. Ghimire, Mr. J. K. Mallik; Students from Acme and Khowpa engineering college: Mr. B. Adhikari, Ms. S. Regmi, Mr. B. Thapa, Mr. P. Amagain, Mr. M. Niraula, and Mr. M. Humagain.**REFERENCES**

A. Nakarmi. (2014). Policy need for deployment of roof top solar PV systems in urban areas of Nepal Center for Energy Studies, Institute of Engineering, TU, Nepal.
CBS. (2011). National Population and Housing Census 2011. Volume 02, NPHC 2011
D. Chianese, D. Pittet, J.N. Shrestha, D. Sharma, A. Zahnd, N. Sanjel, M. Shah and M. Uphadyaya. (2009). Feasibility study on Grid connected PV system in Nepal.

G. R. Pokharel. (2016). Renewable energy, energy mix and energy security in Nepal, Tribhuvan University, Institute of Engineering, Thapathali Campus

G.D. Isreal. Determining sample size. University of Florid, Department of Agricultural Education and Communication, Institute of Food and Agricultural Sciences (IFAS), University of Florida, Gainesville, FL 32611.
Ministry of Industry, Department of Industry Planning, Monitoring & Evaluation Division. (2014). Industrial statistics Fiscal Year 2069/070 (2012/2013). Government of Nepal, Tripureswore, Kathmandu

Nepal Electricity Authority (NEA). (2016). A year in review fiscal year -2015/2016 and 2014/2015, Kathmandu, Nepal
S. M. Nepal. (2014). Freeing the Grid Solar PV System for Kathmandu.
TERAI. (2014). Reaching the sun with roof top solar. New Delhi